# Gene Set Summarization Using Large Language Models

Marcin P. Joachimiak[1], J. Harry Caufield[1], Nomi L. Harris[1], Hyeongsik Kim[2], Christopher J. Mungall[1]

[1]Biosystems Data Science Department, Environmental Genomics and Systems Biology Division, Lawrence Berkeley National Laboratory, 1 Cyclotron Road, Berkeley, CA 94720, USA
[2]Robert Bosch LLC, Sunnyvale, CA 94085, USA

## Abstract

Molecular biologists frequently interpret gene lists derived from high-throughput experiments and computational analysis. This is typically done as a statistical enrichment analysis that measures the over- or under-representation of biological function terms associated with genes or their properties, based on curated assertions from a knowledge base (KB) such as the Gene Ontology (GO). Interpreting gene lists can also be framed as a textual summarization task, enabling Large Language Models (LLMs) to use scientific texts directly and avoid reliance on a KB.

TALISMAN (Terminological ArtificiaL Intelligence SuMmarization of Annotation and Narratives) uses generative AI to perform gene set function summarization as a complement to standard enrichment analysis. This method can use different sources of gene functional information: (1) structured text derived from curated ontological KB annotations, (2) ontology-free narrative gene summaries, or (3) direct retrieval from the model.

We demonstrate that these methods are able to generate plausible and biologically valid summary GO term lists for an input gene set. However, LLM-based approaches are unable to deliver reliable scores or p-values and often return terms that are not statistically significant. Crucially, in our experiments these methods were rarely able to recapitulate the most precise and informative term from standard enrichment analysis. We also observe minor differences depending on prompt input information, with GO term descriptions leading to higher recall but lower precision. However, newer LLM models perform statistically significantly better than the oldest model across all performance metrics, suggesting that future models may lead to further improvements. Overall, the results are nondeterministic, with minor variations in prompt resulting in radically different term lists, true to the stochastic nature of LLMs. Our results show that at this point, LLM-based methods are unsuitable as a replacement for standard term enrichment analysis, however they may provide summarization benefits for implicit knowledge integration across extant but unstandardized knowledge, for large sets of features, and where the amount of information is difficult for humans to process.

# Introduction

Molecular biologists frequently need to interpret the results of experiments or investigations that result in lists of genes. These gene lists are routinely used to infer underlying mechanisms or phenomena. For example, a readout of genes expressed in cancer cells may be used to infer underlying signaling pathways, which in turn can suggest therapeutic approaches. Alternatively, a Genome-Wide Association Survey (GWAS) investigating a trait or disease may reveal correlations between that trait and variants in one or multiple genes.

The standard approach to inferring the underlying mechanism driven by a small set of genes is an open-ended exploratory analysis. Researchers investigate each gene individually in the literature and databases, then synthesize this curated knowledge into a summary and hypothesis. It may even be possible for a researcher to do this based on common knowledge of the genes involved in a pathway. This approach is time-consuming, subjective, and prone to bias, even for small gene sets. For larger gene sets, it is completely infeasible and while some automated approaches can leverage mapping to pathways or other biological classifications (Reimand et al., 2019; Zhao & Rhee, 2023), even in these cases inference of relevant biological patterns still involves human insight to cover cases such as pathways gaps, annotation errors, or functional degeneracy. In practice, researchers typically perform a gene set *enrichment* or *over-representation analysis*, in which curated *ontological annotations* of these genes are extracted and optionally compared against the annotations of the background set. These analyses make use of knowledge bases (KBs) with two components: (1) an ontology, which provides a hierarchical logical organization of gene function descriptors; and (2) gene annotations, which associate genes with these descriptors. The ontology supports the analysis by enabling reasoning to generalize to broader terms and allowing assessment of term information content.

These enrichment analyses are part of the core fabric of molecular biology and biomedical research. The leading system is the Gene Ontology(Gene Ontology Consortium, 2023), which provides ontological annotations of genes across the tree of life using over 40,000 descriptor terms. The GO is one of the most widely cited tools in the life sciences(Duck et al., 2016), and hundreds of tools implement GO enrichment analyses for a range of experimental modalities, from single cell analysis to GWAS. For example, a recent study measured gene expression at the single cell level in multiple cell populations in the human brain vasculature(Garcia et al., 2022). Each population was analyzed using GO, revealing functional roles of different cell subtypes, with implications for conditions involving cerebrovascular injury.

More recently, instruction-based Large Language Models (LLMs) based on the Generative Pre-Trained (GPT) architecture(Brown et al., 2020) have attracted attention due to their highly general abilities on a wide range of text processing tasks, including information extraction, query construction, question answering, and text summarization. Instruction-based LLMs such as GPT-3 and successors are distinguished from the previous generation of models such as BERT(Devlin et al., 2018) and BioBERT(Lee et al., 2019) by their ability to follow instructions in response to a prompt, and the ability to generalize from a small number of examples (few-shot or in-context learning). We have demonstrated that instruction-based LLMs can be used in conjunction with ontologies for KB and ontology extraction tasks(Harry Caufield, Hegde, et al., 2023), potentially as an aid to curation. Others have demonstrated the ability of LLMs to perform tasks such as candidate gene prioritization and selection (Toufiq et al., 2023), annotation of single-cell sequencing data(Hou & Ji, 2023), and generating labels for gene sets(Hu et al., 2023). Closest to our work, Hu et al. performed an evaluation of generated gene set labels compared to the original human-assigned set labels and this work has conceptual overlaps with our approach to evaluate LLMs for gene set summarization.
Here we investigate the ability of LLMs to interpret lists of genes, such as those yielded by gene expression experiments and GWAS. We do this by reframing the task from one of *statistical enrichment* to a *text summarization* task, i.e. taking a larger text and condensing it into salient points. We devised a method that uses LLMs and configurable sources of gene descriptions to perform summarization, taking as input a gene set and producing as output (1) a list of relevant terms, analogous to enriched terms in an over-representation analysis; and (2) a descriptive summary that weaves together the different functions.

We explore three different summarization approaches. The first (which we call “no synopsis”) is purely generative, and relies solely on the massive corpus of documents ingested as training for the GPT model (which can be thought of as the “latent KB” of the model). The second (“narrative synopsis”) makes use of narrative gene summaries, such as those authored by the curators of the RefSeq database(O’Leary et al., 2016). The third (“ontological synopsis”) makes use of controlled textual summaries of GO annotations, such as those provided by the Alliance of Genome Resources (AGR)(Kishore et al., 2020). We evaluate all methods against standard statistical enrichment analysis (see Methods).

# Methods

## TALISMAN: A novel method for gene set summarization using language models

We created a method for summarizing gene sets using LLMs called TALISMAN: Terminological ArtificiaL Intelligence SuMmarization of Annotation and Narratives. This method takes as input a list of N genes $g_1, g_2, \ldots, g_N$ and produces two outputs: (1) a textual summarization of salient features of the gene set, and (2) a list of M ontology terms $t_1, t_2, \ldots, t_M$ from an ontology such as the GO. The method works by generating a structured prompt containing textual summaries of genes from a list of sources. The prompt is also crafted to instruct the model to extract salient features of the gene sets (Fig. 1). The method is intended for LLMs that have been fine-tuned on instruction-following tasks, such as GPT-3.5, GPT-4, as well as open models such as Llama2 (Touvron et al., 2023).

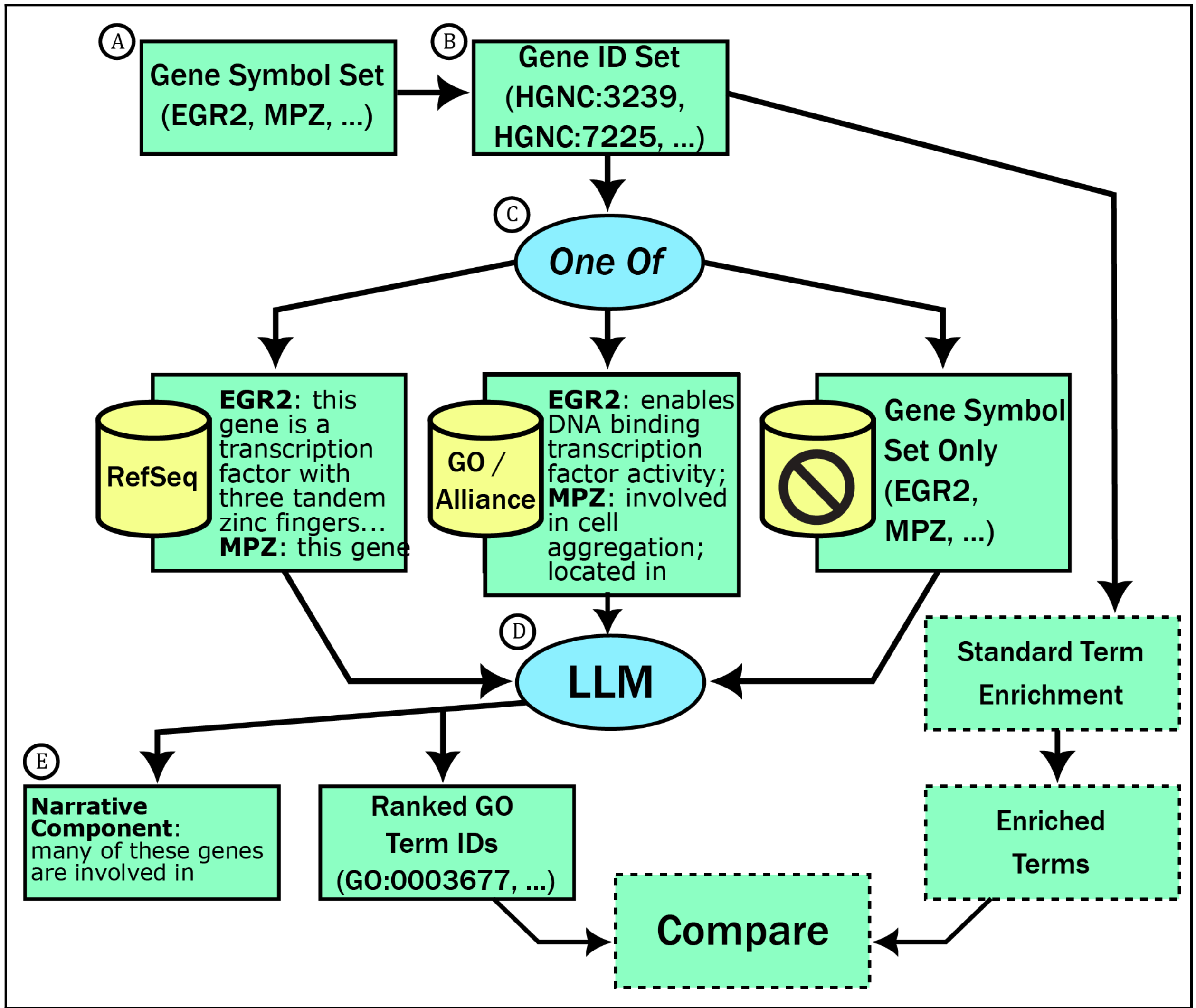

**Figure 1**: The TALISMAN workflow. A) User selects a set of gene symbols. B) Gene symbols are parsed to gene identifiers. C) A textual description of the genes is generated using one of three approaches, either narrative summaries from RefSeq, controlled natural language derived from the Gene Ontology and Alliance of Gene Resources collections, or gene symbols only (i.e., no additional descriptions). D) The text from the previous step is used to construct a completion prompt for an LLM. E) The results of the prompt are parsed into a narrative component and a ranked list of terms. As part of the evaluation (boxes with dashed outline) we compared TALISMAN with terms from standard enrichment.

## Structured Prompt Generation

For each gene ID in the gene set, we query the Alliance of Genome Resources API (Alliance of Genome Resources Consortium, 2020) to retrieve (i) the gene symbol; (ii) a narrative gene description, aggregated from RefSeq; and (iii) automated gene descriptions (Kishore et al., 2020). Note that automated gene descriptions are in fact derived from curated ontological GO annotations; here “automated” refers to the ontology-to-text process rather than the process of generating the ontological annotations in the first place.

For each gene we generate a *description* that is a concatenation of the gene symbol and the description separated by a colon character. For the narrative method, we use the narrative gene description, and for the ontological method we use the ontology term summaries. For the generative approach, we only provide the gene symbols.

We then generate a prompt using the Jinja template system(Ronacher, 2008) with a standard template incorporating the gene description lines, with two template variables, “taxon”, and “gene descriptions” The input to Jinja is as follows:

> *I will give you a list of {{ taxon }} genes together with descriptions of their functions.*
> *Perform a term enrichment test on these genes.*
> *i.e. tell me what the commonalities are in their function.*
> *Make use of classification hierarchies when you do this.*
> *Only report gene functions in common, not diseases.*
> *e.g. if gene1 is involved in "toe bone growth" and gene2 is involved in "finger morphogenesis"*
> *then the term "digit development" would be enriched as represented by gene1 and gene2.*
> *Only include terms that are statistically over-represented.*
> *Also include a hypothesis of the underlying biological mechanism or pathway.*
>
> *Provide results in the format*
>
> *Summary: <high level summary>*
> *Mechanism: <mechanism>*
> *Enriched Terms: <term1>; <term2>; <term3>*
>
> *For the list of terms, be sure to use a semicolon separator, and do not number the list.*
> *Always put the list of terms last, after mechanism, summary, or hypotheses.*
>
> *Here are the gene summaries:*
> *{{ GENE_DESCRIPTIONS }}*

Note that the header includes an in-context directive specifically instructing the model to generalize over the gene sets, including providing an example of how to do this.

## Token Length Limits and Gene Description Truncation

One of the current limitations of LLMs is the number of tokens (roughly, the number of words) that can be provided as both input and output. If gene lists are large, or the textual summaries of the genes are long, then the prompt will exceed the maximum token length (currently 4k for GPT-3.x models, and 8k or 32k for GPT-4).

In order to accommodate these limits in different models we truncate the length of each gene description proportional to the total number of tokens relative to maximum token length. We truncate from the end of the string, on the assumption that the text at the beginning is more informative.

Note this can result in substantial information loss, proportional to the number of genes in the input gene set. We record this as the truncation factor (TF); a TF of 1.0 reflects that the prompt was generated without truncation, while a TF of 0.25 indicates that only 25% of the original text could be used.

## Prompt Completion and Payload Parsing

Generated prompts are fed to the model via the OpenAI API. We use the default configuration, with the lowest "temperature" (creativity) setting (i.e. maximizing determinism). Results are cached to avoid expensive recomputation.

Our approach to prompt completion parsing reuses the method described in our SPIRES manuscript(Harry Caufield, Hegde, et al., 2023), in which the resulting functional terms are grounded (mapped to terms) in the Gene Ontology using the Ontology Access Kit (OAK)(Creators Chris Mungall1 Harshad1 Patrick Kalita1 Charles Tapley Hoyt2 Sujay Patil1 marcin p. joachimiak1 Joe Flack3 David Linke4 Nomi Harris1 Sierra Moxon5 Kim Rutherford6 Nico Matentzoglu7 Deepak8 Harry Caufield1 Vinícius de Souza Glass9 Jules Jacobsen10 Justin Reese11 Manuel Lera Ramirez Shawn Tan Show affiliations 1. Lawrence Berkeley National Laboratory 2. Harvard Medical School 3. @jhu-bids 4. Leibniz-Institut für Katalyse e. V. (LIKAT) 5. LBNL 6. Uni of Cambridge / @PomBase 7. semanticly Ltd 8. SIB Swiss Institute of Bioinformatics 9. @det-lab @tis-lab @monarch-initiative 10. Queen Mary University of London 11. Lawrence Berkeley National Lab, 2023) annotate functionality.

Note that our prompt asks for separate sections in the payload: a high level narrative summary plus a list of terms. The narrative summary is not parsed by TALISMAN and is presented to the user as-is. The string with the list of terms is split, and the resulting list is fed through the OAK annotator. This step assumes that the model yields descriptors that conform to the terminology of GO, using either the primary label or the synonym.

Our prompt explicitly avoids asking for GO identifiers or any other form of identifier. This is because we and others (Bombieri et al., 2024; Harry Caufield, Hegde, et al., 2023; Pal et al., 2023) have observed that GPT models frequently hallucinate "likely seeming" numeric identifiers, consistent with the design of sequential generation methods which are not trained on absolute truth and strive to maintain a variety of output.

## Implementation

The TALISMAN code is available at and the associated data for this study is here [https://github.com/monarch-initiative/talisman-paper](https://github.com/monarch-initiative/talisman-paper).

TALISMAN is implemented in Python (https://github.com/monarch-initiative/talisman) and includes dependencies on the recently published OntoGPT package (https://github.com/monarch-initiative/ontogpt)(Harry Caufield, Hegde, et al., 2023).

Access to a GPT model via an API such as the OpenAI API is required. However, for evaluation purposes, it is possible to use our cached completions. TALISMAN is agnostic to the exact instruction-tuned LLM assuming training on large corpuses which include GO terms, annotations, and gene function descriptions. Due to differences in reinforcement learning from human feedback (RLHF) the TALISMAN prompt may need to be optimized to generate results in the expected format.

Prompt completions are cached in a local sqlite3 database to avoid incurring charges by repeated requests of the same text. There is an interactive TALISMAN mode that bypasses API access and asks the user to copy the prompt into the web ChatGPT interface, and then copy the results back.

We provide both a command line interface and a web application interface for TALISMAN. The web application interface makes use of the Streamlit framework, and currently must be executed locally. The web application UI is shown in Figure 2.

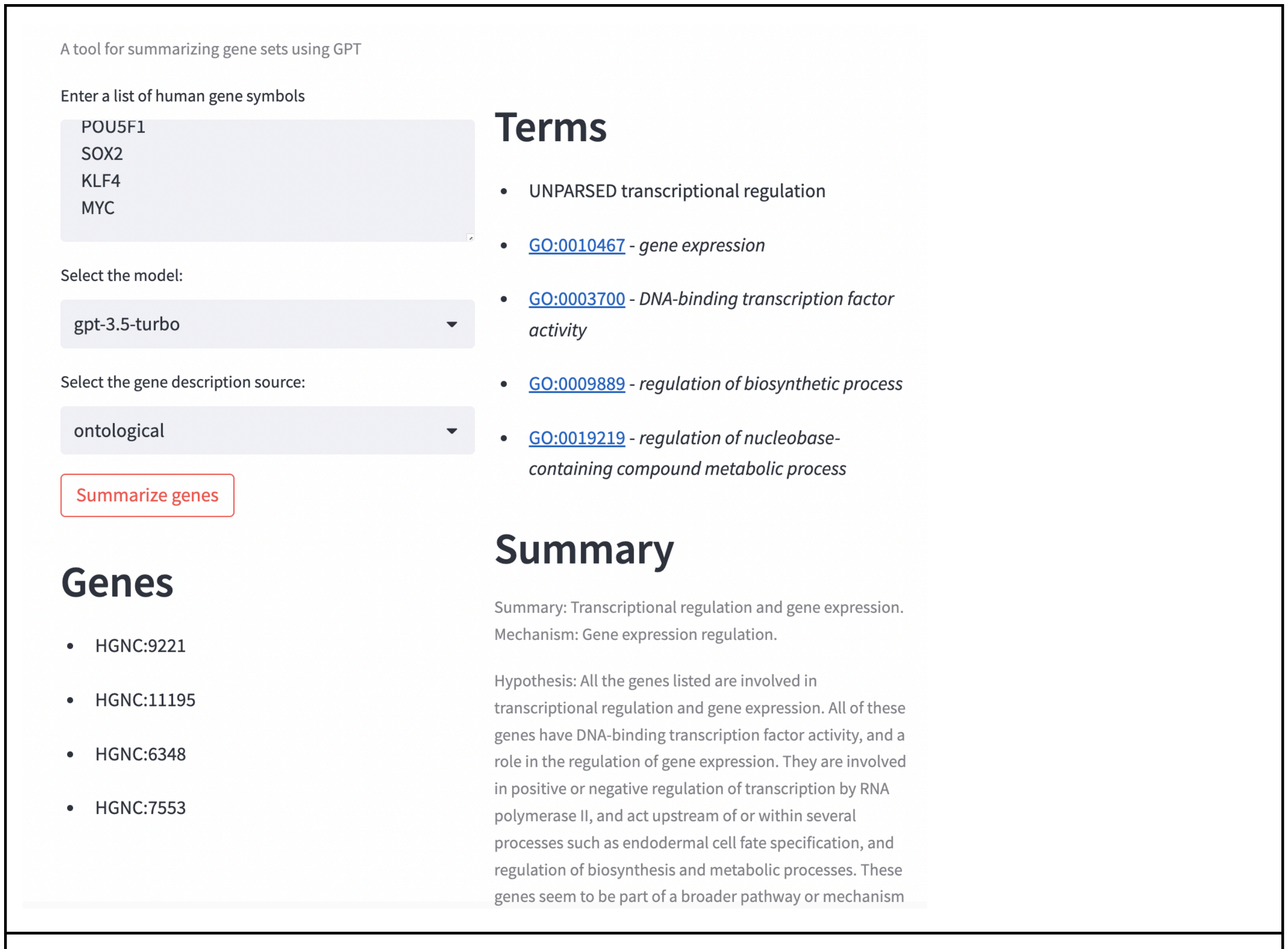

**Figure 2**: UI for TALISMAN web application

## Evaluation

There is no single agreed-upon approach to benchmarking enrichment analysis algorithms (Ballouz et al., 2017). For this study, we collected 70 human gene sets for evaluation, ranging in size from 3 to 200 genes, from multiple sources including MSigDB(Dolgalev, n.d.), GeneWeaver (Baker et al., 2016), Human Phenotype Ontology Annotations(Köhler et al., 2021), disease to gene relationships from the Monarch Initiative(Putman et al., 2024), sample of large biclusters [REF] based on RNAseqDB data (Wang et al., 2018), and, as a baseline, gene sets corresponding to existing terms in the GO. For the main evaluation all gene sets consist of human genes.

For each gene set, we generated an additional perturbed gene set simulating noise, where we dropped out 10% of genes and inserted random genes as replacements.

For each gene set and perturbed gene set we ran the three TALISMAN methods (see Figure 1) with three different models, GPT-3.0 (also known as *text-davinci-003), GPT-3.5*, and GPT-4.

We compare the results of TALISMAN with standard statistical gene set enrichment implemented in OAK, using hypergeometric tests and Bonferroni correction.

Due to the nature of gene set enrichment, we expect the resulting enriched terms to be different. In particular, different enrichment tools may choose terms at different levels in the hierarchy, each representing valid perspectives. Additionally, the gene set summarization method doesn't return *p value* calculations, which makes it harder to compare. We therefore employed a method that compared enrichment results using different parameters and cutoffs.

In order to compare, we first filtered the results of standard enrichment, taking only the top *n* results (as most *n*) for a given p-value cutoff of *p*. For each term $t_{ann}$ in the standard enrichment results, if there exists a predicted term $t_{pred}$ that is equal to or is an ancestor or descendant of t, it counts as a true positive. Other terms in the predicted set that are unaccounted for count as false positives, and other terms in the standard enrichment results that are unaccounted for count as false negatives.

We score these outcomes for different values of *n* and *p* (***Sup. Fig. X***). For the case where *n*=1, this corresponds to checking whether the top standard enrichment result term is recapitulated.

We also calculated a simplified metric *has hit*, which is 1 or 0 depending on whether the predicted terms included any term from the top *n*. In the case where *n*=1 we call this *has top hit,* as it measures whether the best ranking term from standard enrichment is found at all in the predicted set.

Distributions of precision, recall, and F1 score values were compared between pairs of different model results using the exact Mann-Whitney test.

The results of the evaluation are available via Zenodo (Joachimiak, 2023) and can be viewed as a Jupyter Notebook (https://github.com/monarch-initiative/enrichgpt-results).

# Results

## Using GO annotations as a source best recapitulates gold standard annotations top hit

We curated 70 gene sets and ran all methods on each gene set plus a perturbed copy of each gene set (see above). Next, we tested three different GPT models: GPT-3.0, its successor *GPT-3.5*, as well as GPT-4.0. For each model, we tested the three sources of gene descriptions: ontological synopses (GO), narrative synopses (RefSeq), and no synopsis (None). We evaluated the results of all methods across all gene sets. We deposited these results in a Zenodo-archived GitHub repository (https://github.com/monarch-initiative/enrichgpt-results/).

| Source | Model | Has Top Hit |
| --- | --- | --- |
| GO | GPT-4.0 | **0.865** ± 0.343 |
| NONE | GPT-3.5 | 0.812 ± 0.391 |
| NONE | GPT-4.0 | 0.799 ± 0.402 |
| RefSeq | GPT-4.0 | 0.757 ± 0.43 |
| RefSeq | GPT-3.5 | 0.74 ± 0.44 |
| GO | GPT-3.5 | 0.698 ± 0.46 |
| GO | GPT-3.0 | 0.615 ± 0.488 |
| NONE | GPT-3.0 | 0.559 ± 0.497 |
| RefSeq | GPT-3.0 | 0.517 ± 0.501 |

**Table 1**: Proportion of runs in which the top hit (i.e most enriched term) from standard enrichment is recapitulated (either directly or through ancestry) using LLMs. These results are for p-value < 0.05, top 10 hits, and with ontological closure.

We analyzed the runs across all gene sets, computing the proportion of runs in which the best ranking term from standard enrichment was present in the LLM provided gene set for each description source and model combination (Table 1). The best performance is achieved by using GO as the source of gene descriptions in combination with the GPT-4.0 model. Surprisingly GPT3.5 with None is second in this metric, however, the same model with GO term descriptions is significantly worse than the best result. GPT-3.0 underperforms in all cases independent of the description source. The best performing methods also exhibit a lower standard deviation (SD) meaning that the higher performance is consistent across different gene sets.

## Comparison of method and source combinations reveals enrichment performance trends

We compared the gene set summarization behavior of different LLM method, model, and description source combinations using standard model performance metrics. In this assessment we used the reference gene sets as true answers, considered the top 10 most significantly enriched genes from standard enrichment, included ontology closure terms, and did not allow for any new, potentially correct information from the LLMs (see Methods). The mean precision, recall, and F1 scores were derived across all the gene sets for each method, model, and description source combination. Results for GPT-4.0 (Fig. 3) showed a higher recall and F1 score when using GO term descriptions, compared to no synopsis and RefSeq gene descriptions. Recall followed a similar trend with highest recall observed when using GO term descriptions, followed by no synopsis and then RefSeq descriptions. However, the precision trend was the reverse: no synopsis gave the highest recall, followed by RefSeq gene and then GO term descriptions. Thus we observe a tradeoff between recall and precision, with GO information giving the highest recall and no synopsis giving the highest precision. The former result suggests that including GO information leads to reporting of more significantly enriched GO terms but this comes at the expense of also reporting additional non-significant GO terms.

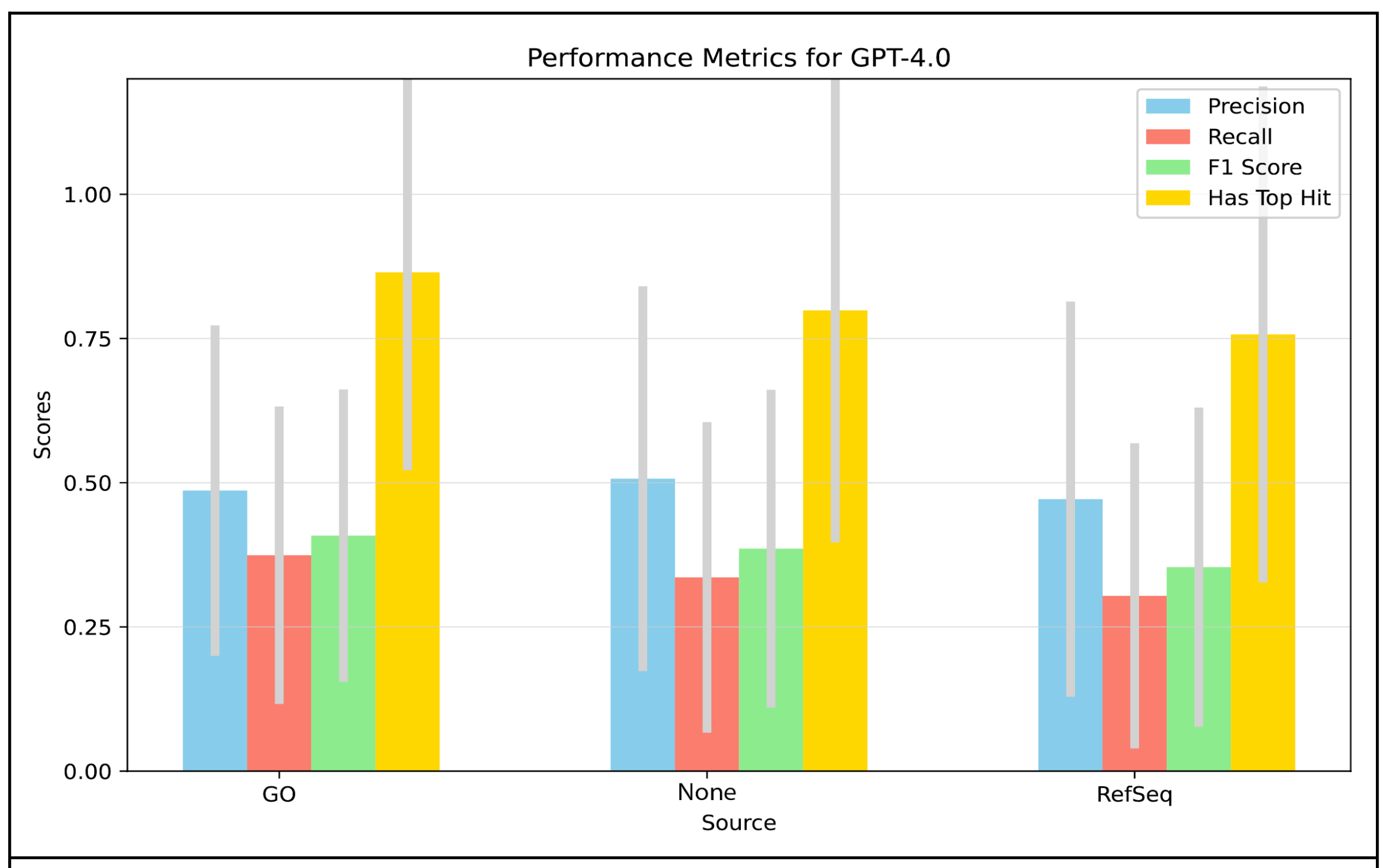

**Figure 3.** Bar chart comparing gene set enrichment performance of description source combinations for GPT-4.0. These results are for p-value < 0.05, top 10 hits, and with ontological closure.

When including other GPT models (Table 2), GPT-3.0 and 3.5, additional trends were observed. The most sophisticated model, GPT-4, with ontology term summaries (“GO”) has the highest recall and fraction of results with a top enriched term. However, it also has lower precision and F1 than the GPT-3.5 model with only gene names. On the other hand when considering ontology summaries, each model iteration showed a clear increasing trend except for lower precision for GPT-4.0 versus 3.5 versus 3.0. Surprisingly, using no synopsis often gave better performance than using RefSeq or GO descriptions. The best performance was with GPT-3.5 and no synopsis data (GPT-3.5-NONE in Table 2), with the highest precision and F1, and as well as GPT-4.0 with GO term descriptions (GPT-4.0-GO), with highest recall and proportion of top hits. This result may be a reflection of the model training, with gene symbols alone (“None”) giving more accurate information than additionally providing concise and curated biological knowledge such as RefSeq or GO descriptions. We also observed that including RefSeq descriptions gave better performance than GO term descriptions for GPT-3.5, but for GPT-4.0 this was only true for precision but not recall and F1. Tradeoffs of precision versus recall apply in different settings, for example during exploratory analysis where higher recall is desired.

| Source | Model | Precision | Recall | F1 score | Has Top Hit |
|---|---|---|---|---|---|
| GO | GPT-3.0 | 0.252 ± 0.266 | 0.2 ± 0.225 | 0.208 ± 0.22 | 0.615 ± 0.488 |
| GO | GPT-3.5 | 0.467 ± 0.371 | 0.265 ± 0.256 | 0.318 ± 0.278 | 0.698 ± 0.46 |
| GO | GPT-4.0 | 0.486 ± 0.286 | **0.374** ± 0.258 | 0.408 ± 0.253 | **0.865** ± 0.343 |
| NONE | GPT-3.0 | 0.339 ± 0.363 | 0.171 ± 0.21 | 0.211 ± 0.238 | 0.559 ± 0.497 |
| NONE | GPT-3.5 | **0.543** ± 0.343 | 0.371 ± 0.278 | **0.42** ± 0.28 | 0.812 ± 0.391 |
| NONE | GPT-4.0 | 0.507 ± 0.333 | 0.336 ± 0.269 | 0.386 ± 0.275 | 0.799 ± 0.402 |
| RefSeq | GPT-3.0 | 0.261 ± 0.308 | 0.172 ± 0.225 | 0.196 ± 0.24 | 0.517 ± 0.501 |
| RefSeq | GPT-3.5 | 0.517 ± 0.38 | 0.31 ± 0.276 | 0.368 ± 0.295 | 0.74 ± 0.44 |
| RefSeq | GPT-4.0 | 0.471 ± 0.342 | 0.304 ± 0.264 | 0.354 ± 0.277 | 0.757 ± 0.43 |

**Table 2.** Table comparing gene set enrichment performance of description source combinations across GPT models. These results are for p-value < 0.05, top 10 hits, and with ontological closure.

When averaging over all cases and standard enrichment p-value cutoffs models (**Table 3**), additional trends were observed. The most sophisticated model has the highest recall, F1, and fraction of results with a top enriched term but with somewhat lower precision than the GPT-3.5 model. Since the GPT-3.5-None improvement is seen at the highest p-value cutoff but not all averaged results, this suggests that the precision difference becomes smaller as less significant terms are included.

| Model | Precision | Recall | F1 score | Has Top Hit |
|---|---|---|---|---|
| GPT-3.0 | 0.263 ± 0.368 | 0.108 ± 0.227 | 0.121 ± 0.231 | 0.438 ± 0.496 |
| GPT-3.5 | **0.423** ± 0.418 | 0.188 ± 0.293 | 0.216 ± 0.3 | 0.598 ± 0.49 |
| GPT-4.0 | 0.414 ± 0.398 | **0.198** ± 0.295 | **0.223** ± 0.296 | **0.64** ± 0.48 |

**Table 3**: Comparison of different models averaging across all gene description sources and all evaluation parameters. GPT-4.0 performs best on all metrics other than precision where it is narrowly beaten by GPT-3.5. The older GPT-3.0 model consistently performs worse. These results are for all evaluation parameters.

To formalize the model comparisons, we performed pairwise model statistical tests for precision, recall, and F1 score value distributions (Figure 4). The analysis revealed a consistent result with all GPT-3.5 and 4.0 model results being significantly better than any GPT-3.0 results. This confirms advances in model architecture and training, with increases in both model complexity (e.g. number of parameters) and input training data. The two best models were GPT-3.5 with no input data and GPT-4.0 with GO term descriptions, which were significantly better than all other combinations (except GPT-3.5-None versus GPT-4.0-None). Overall, no model was significantly better than either GPT-3.5 or GPT-4.0 with GO term descriptions.

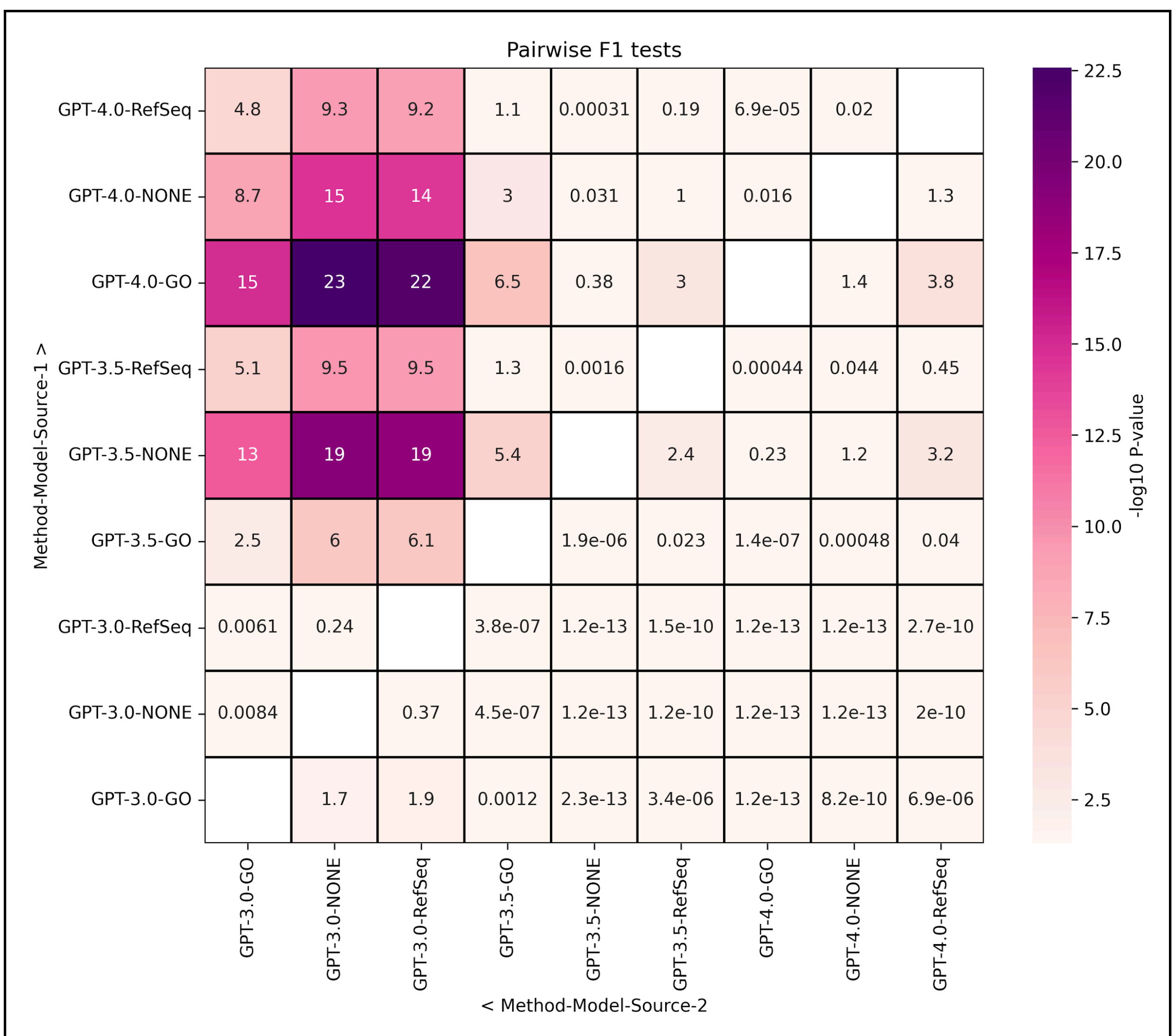

**Figure 4** Heatmap of -$\log_{10}$(p-value) statistical test results for pairwise model comparisons of GO term enrichment F1 scores. Distributions of values for method, model and description source combinations (y-axis) were compared using a one-sided nonparametric exact test. Values that are p-value < 0.05, corresponding to the value of -$\log_{10}$(p-value)=1.3, are in darker shades.

Extending this analysis to precision and recall values (Sup. Info.), we also observe that all GPT-4.0 and 3.5 results are significantly better than 3.0. However, no GPT-4.0 model was significantly better than any other GPT-4.0 or 3.5 model. As in the F1-score results, GPT-3.5-NONE precision and recall were significantly better than GPT-3.5-GO, as well as GPT-4.0 with GO and RefSeq. Notably, GPT-3.0-NONE was significantly better in precision and recall than GPT-3.0 with GO or RefSeq, just as observed for GPT-3.5. This result was not observed for the GPT-4.0 model suggesting something qualitatively different in how additional information such as GO term descriptions or RefSeq narratives are functionally used by a newer model.

## Gene set summaries are biologically plausible in a way that disguises limitations

To gain a better understanding of how AI-based gene set summarization differs from standard statistical enrichment, we performed a qualitative assessment of the results of GPT summary derived term lists. When examined in isolation, these term lists were largely biologically plausible, valid (i.e. at least one gene that had the indicated function) across all models, and regardless of source of gene descriptions. However, when the results for a given gene set were compared across methods or compared to the gold standard statistical-ontological enrichment it was revealed that results are often close to standard enrichment results albeit XXXX.

This can be seen in Figure 6, which shows superimposed results (GPT-3.5 only) for genes associated with the Human Phenotype Ontology term “Sensory ataxia” (HP:0010871; EGR2 NAGLU GPI DNAJC3 SH3TC2 TWNK PIEZO2 FLVCR1 MPZ PRX PMP22 KPNA3 POLG RNF170 AARS1). We selected this gene set intentionally as an “easy” set with a clear underlying mechanism, to see what a good TALISMAN result might look like. Genes implicated in Mendelian diseases such as sensory ataxia are more likely to be studied and annotated. This particular phenotype of sensory ataxia has been well studied, with a large literature on underlying pathophysiological mechanisms(Lopriore et al., 2022).

Standard GO over-representation on this gene set yields “myelination” and “Schwann cell differentiation” as top hits (lowest p-value). Figure 6 shows all terms found by TALISMAN using different gene description sources, compared against standard enrichment. The significant and gene-set relevant GO term “myelination” was found when using either ontologies as gene description or providing gene descriptions; however, when using narrative gene descriptions as a source, the string “myelin sheath maintenance” is returned, which essentially means the same thing, but automated methods do not ground this term and hence do not reveal the equivalence. Only the narrative based method found “mitochondrial DNA replication”. None of the GPT methods detected “Schwann cell differentiation”.

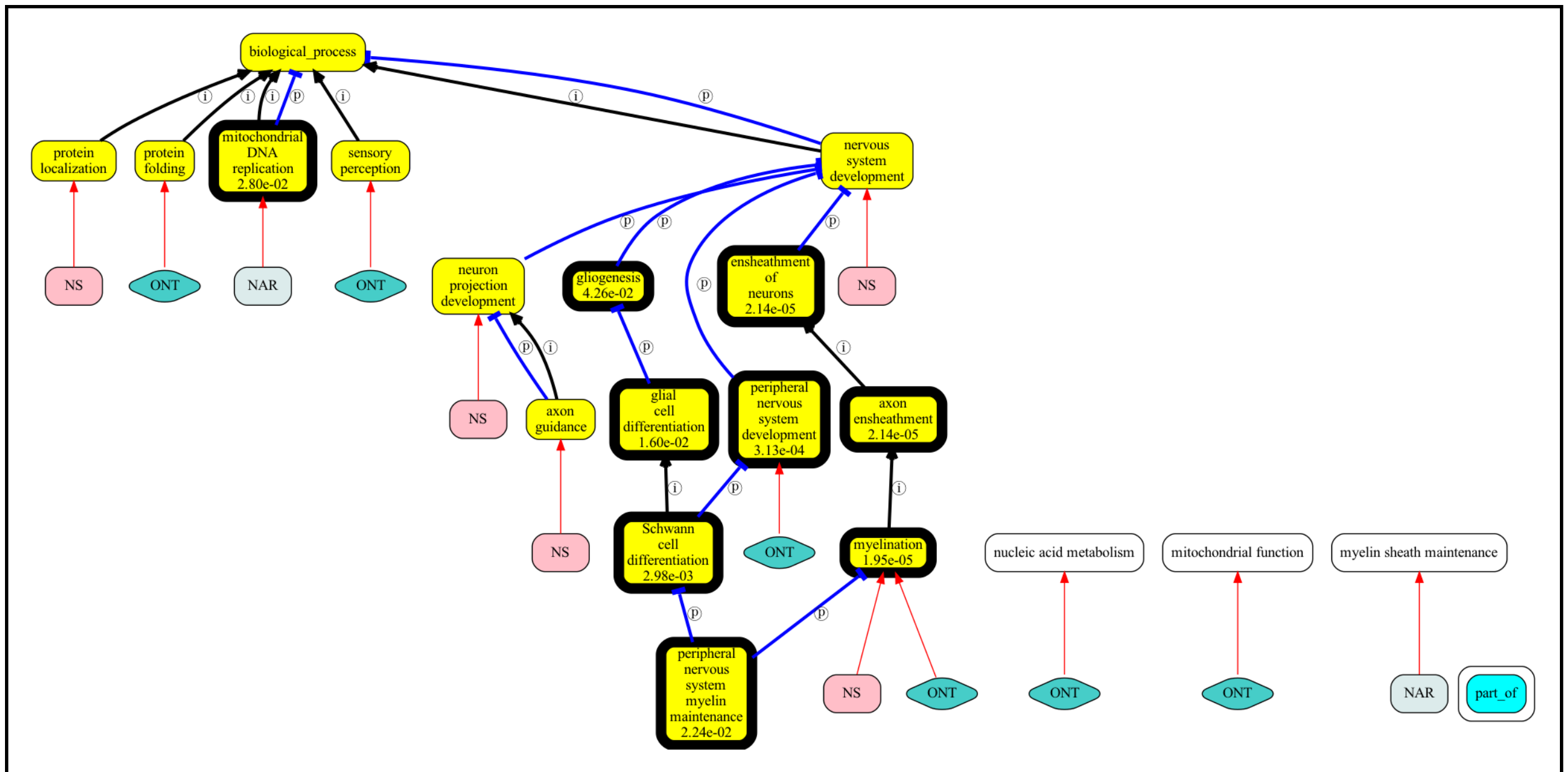

**Figure 6**: Superimposed results for summarization of sensory ataxia gene set (genes annotated to “Sensory ataxia”; HP:0010871). GO terms are in yellow boxes; a bold border indicates significance (p-values shown in the text box). Terms in white boxes are those that could not be grounded to a GO term identifier. ONT=ontological synopsis, NS=no synopsis, NAR=narrative synopsis.

## GPT returns highly variable answers across different runs

To investigate the stability of LLM results we performed two runs for each model-method combination, where on the second run we made an insignificant syntactic change to the prompt (changing the end marker from 3 hashes (###) to 3 equals symbols (===)). We then measured the Jaccard similarity of the term sets of each run (counting terms directly rather than using the ontology hierarchy). There was a very low level of consistency across runs, with the most consistent being GPT-3.5 with no synopses. Consistency was nearly twice as high for GPT-4.0 vs. GPT-3.5 and almost 6-fold higher relative to GPT-3.0.

| | | Mean Jaccard | Jaccard std | Jaccard min | Jaccard max |
|---|---|---|---|---|---|
| **model** | **method** | | | | |
| GPT-3.5 | RefSeq | 0.152 | 0.143 | **0** | 0.75 |
| | None | 0.123 | 0.129 | **0** | 0.5 |
| | GO | 0.16 | **0.185** | **0** | 0.8 |
| GPT-3.0 | RefSeq | 0.061 | 0.07 | **0** | 0.333 |
| | None | 0.038 | 0.052 | **0** | 0.25 |
| | GO | 0.084 | 0.095 | **0** | 0.5 |

| | RefSeq | 0.179 | 0.144 | **0** | 0.833 |
|---|---|---|---|---|---|
| | **None** | **0.219** | 0.158 | **0** | **1** |
| **GPT-4.0** | **GO** | 0.172 | 0.147 | **0** | 0.889 |

**Table 3**: Summary statistics for Jaccard similarity of term lists (N=142) when prompt is modified across all gene sets. Higher mean values are more favorable.

## Generated narrative summaries are plausible but non-deterministic

We also examined the textual summaries produced by the GPT-3.5 model with the three sources of gene descriptions. An example is provided in Table 4, showing the results for the sensory ataxia gene set.

| **Source** | **Summary** | **Mechanism** |
|---|---|---|
| Ontological synopsis (GPT-4.0) | The provided genes are mainly involved in processes related to the nervous system, peripheral nerve function, and cellular maintenance functions. | These genes may contribute to the biological processes related to the nervous system development, cellular response regulation, and transportation of molecules within cells, interacting in various pathways. |
| Narrative synopsis (GPT-4.0) | Majority of the genes are associated with neuropathic conditions and myelin-related processes in the peripheral nervous system. | The underlying biological mechanism may be related to the formation, maintenance, and function of the myelin sheath in the peripheral nervous system and the regulation of cellular pathways that impact neuronal survival and function. |
| No synopsis (GPT-4.0) | Enriched terms associated with the given list of genes are mostly involved in the development and maintenance of the nervous system, cellular response, and transport processes. | These genes may contribute to the biological processes related to the nervous system development, cellular response regulation, and transportation of molecules within cells, interacting in various pathways. |

**Table 4**: Textual summary of the sensory ataxia gene set descriptions using the 3 different approaches.

Overall, the qualitative summaries of the gene sets appear plausible, although inconsistent in being able to yield the most significant term. We note that sentences using 'involved in' versus associated with' have high semantic similarity (cosine similarity = 0.98 for the first and third sentence in Table 4) and close natural language meaning. In addition the statement 'enriched in' does imply that this term or genes are statistically enriched in this case.

## Hallucinations are rare to non-existent when summarizing human gene sets

A common problem with LLMs is the tendency to hallucinate(Ji et al., 2022). Previously we have observed that hallucinations are less problematic for knowledge-oriented *in-context* tasks (Harry Caufield, Hegde, et al., 2023). Here we evaluate the extent to which LLMs hallucinate on a constrained gene set summarization task.

We took all GPT-3.5 model results, and aggregated all unannotated terms for all results. These represent potential hallucinations (i.e. where the model fabricated a term for a gene set). We examined each instance and evaluated whether it was a reasonably valid term for that gene set. Here the criteria for reasonable validity was whether the term was descriptive for any gene in that gene set. We were unable to detect any true hallucinations - every term reported by GPT was in some way reasonably valid even if it did not meet the bar for GO annotation. These unannotated cases fell into three different categories:

- *Use of a term that has been obsoleted in the GO*: In this case, the model is likely recalling an GO annotation to an obsolete term.
- *Regulation vs involved in*: for example, the Ehler Danlos Syndrome gene set summarization includes the term *regulation of collagen metabolism*; the actual GO annotation was to a similar term *collagen metabolism*.
- *Alternate perspective*: a gene is annotated to a closely related term where the categorization is debatable. For example, the GPT3.5 model included the term “glucose transmembrane transporter activity” in the summary for the hallmark glycolysis gene set when given narrative gene set descriptions. Surprisingly, none of the GO annotations for any of the genes in the gene set included this term or a descendant of this term. However, one term in the gene set, SLC37A4, is annotated to glucose-6-phosphate transmembrane transporter. Formally this is not a glucose transporter, as glucose-6-phosphate is a derivative of glucose. However, if GO were to make use of the ChEBI has-function-parent relationship when classifying GO terms then this gene would be classified as a glucose transmembrane transporter.

Although we were unable to detect any true hallucinations, it should be noted that many of the terms given in summarizations still fail to be statistically significant, as observed in the precision data (Table 2).

Our hallucination analysis did not extend to the non-GO term summarizations. We observed that in some cases, these summaries included reports of p-values (even though we did not specifically ask for these), and while these looked plausible, they were in fact fabricated.

We also conducted experiments in which we explicitly asked for p-values to be included in the results, and as expected, these looked plausible but were in fact also fabricated. Thus when the requested task falls within general LLM capabilities (text summarization), hallucinations seem to be avoided, but when a request is made for something likely outside its capabilities (calculation of a statistical test), it will hallucinate rather than giving an expected response that this task falls outside of the models’ designed capabilities.

We also conducted experiments where we swapped in gene descriptions from random other genes, to test whether the model was relying on gene symbols and its own latent KB of those genes, rather than the in-context information. For example, when running an analysis over the gene set for *canonical glycolysis*, we swapped out each gene description for a random gene description from a completely different gene set such as endocytosis. If the LLM were making use of its latent KB, then we might expect that the summary terms would still yield glycolysis terms, based on what the LLM "thinks" the genes do. In fact, regardless of whether the source was ontological synopses or narrative descriptions, the model used the descriptions, and summarized these, ignoring the gene symbols.

# Discussion

Thus GPT-3.5 offers a summarization advantage since significantly enriched terms are a larger proportion of the results.

## Limitations of approach

[potential bias for GPT4 from adding GO or RefSeq input relative to just the gene symbols]
[potential search analogy of standardizing gene symbols vs text description]

We have developed and evaluated a method that performs gene set summarization using language models and configurable sources of gene synopses. While this method has some similarities to standard methods of gene set analysis, it is inherently more limited. Some of these limitations may be due to our own method, while some may be inherent in the use of language models:

- *No background sets*. Providing background sets of genes to estimate the background distribution of function representation is crucial for accurate interpretation of results when not all genes were assayed [REF]. Providing descriptions of genes in the background set is challenging for LLMs due to constraints on the number of tokens that can be passed in a single prompt. Even if no gene descriptions are provided and we are making use of the LLM latent knowledge, the number of gene symbols may be too large. In future as newer models and techniques reduce token constraints it may be feasible to incorporate background genes.
- *Lack of statistics*. Standard methods of interpreting gene sets provide some statistical interpretation of the results, whether this is a p-value, or a probability in the case of model-based methods(Bauer et al., 2010). In contrast, language model based approaches rely on patterns in language. Although some have claimed that mathematical reasoning is an emergent ability of LLMs (Wei et al., 2022), we were unable to find a purely LLM-based way to generate reliable, meaningful statistics for results, although this may change in the future. Of course, it is possible to implement a hybrid approach whereby the LLM hands terms off to a dedicated engine that implements the calculation, but this would only be possible for ontological annotation sources, at which point there is no real benefit to using an LLM.

- *Inherent non-determinism.* A current feature of LLMs is that output is highly non-deterministic, with minor variations in prompt resulting in sometimes massively different outputs. For a text summarization task this is not necessarily a problem, as there are many equally valid ways to summarize a task. But this becomes problematic when we try to apply summarization to interpreting scientific results, where we want to reduce arbitrariness and increase repeatability. One possibility here is to run the model multiple times and statistically aggregate the results. We did not attempt to evaluate this here, in part due to the costs the repeated runs would incur, but this may be a promising avenue for future research.
- *Inputs are unordered gene sets, not ranked lists*. Our method takes as input an unordered set of genes, similar to standard over-representation analysis. Many enrichment tools such as the PANTHER enrichment tool used by the GO Consortium(Mi et al., 2019) allow for rankings within the gene sets, applying the appropriate statistical test. We did not investigate the ability of LLMs to make use of ranked inputs. Including some kind of qualitative weights may be successful but we believe that as stated above using the appropriate statistical measure is likely outside current capabilities

Additionally, our analysis has certain limitations. Evaluating and comparing gene set enrichment methods is challenging due to a lack of gold standards and agreed upon metrics. Previous approaches to evaluation include calculating mutual coverage of gene sets(Hung et al., 2012). We include a mutual coverage Jaccard score in our full Jupyter notebook analysis. However, this is not a good method for evaluating text summarization, which is a different task. With a standard enrichment analysis, significance scores can be calculated for all terms in an ontology, but for text summarization the model selects only a small subset of relevant terms.

## Language models are not a good replacement for manual curation

Gene set enrichment and over-representation analyses rely on high-quality curated KBs such as the GO or Reactome. The use of AI and massive LLMs may seem like an opportunity to bypass curation and use information either from selected textual summaries or from a massive corpus of training data. However, this would be a serious mistake. First it is necessary to acknowledge that these LLMs are almost certainly making heavy use of the curated content of these KBs. Information from the GO is replicated in multiple places, from encyclopedic resources such as Wikipedia to major genomics knowledge portals such as the UniProt and NCBI Gene interfaces, and GO enrichment analyses are commonplace in the literature. Annotations are also frequently stored in repositories such as GitHub, which are included in LLM training sets. The power of LLMs to make use of this information seemingly intelligently is indeed remarkable, but this is all derived from highly curated content. This content needs to be constantly updated in light of new scientific knowledge, otherwise the quality of gene set enrichment results decreases significantly(Green & Karp, 2006; Tomczak et al., 2018; Wadi et al., 2016). Furthermore, our results show that when token length is controlled for, the best results are obtained using only gene symbols or with textual representations of ontological annotations; in this scenario, the LLM is essentially regurgitating existing annotations and does not provide any shortcut to curation. Using LLMs as a justification for bypassing curation would be severely misguided, would result in worse results over time, and is fundamentally misguided as the cost of curation is minimal compared with the costs of performing the underlying experiments, with curation costs accounting for less than 0.01% of the whole(Karp, 2016).

Furthermore, we were unable to get LLM approaches to perform the same kinds of ontology-based generalizations we see with standard enrichment analyses, resulting overall in lower precision and key informative terms being missed in results. Additionally, results are highly non-deterministic, with minor prompt variations resulting in different term lists each run. Thus while using a LLM may on the surface at first seem to deliver relevant and plausible results, the user may be unaware that the results are an arbitrary subset of possible results, and that they may be missing crucial information.

## Future Directions

Our methods described here employ effectively what could be considered zero-shot learning, with a small in-context example of how to generalize in a similar manner to the ontological generalization employed in standard term enrichment. It is possible that fine tuning could improve the ability to generalize sets of terms, or even improve the relevancy and significance of these terms.

Our methods did not make use of the conversational abilities of LLMs, as exhibited by ChatGPT. The user has no opportunity to refine responses, or to interrogate results in finer grained detail. We envision future possibilities in which the user is able to enter a dialog, with LLM wrappers able to transparently interact with multiple different biological KBs as exhibited in the GeneGPT system(Jin et al., 2023).

Our approach is currently limited to enrichment using GO terms. Other annotation systems can be used in enrichment analyses to reveal other salient aspects of the genes involved - for example, gene expression using an anatomy ontology(Bastian et al., 2021), or pathway database annotations(Fabregat et al., 2018). In future we will explore gene set summarization methods for these other kinds of annotation systems, as well as unifying methods that can synthesize across multiple knowledge sources, such as found in the Monarch Knowledge Graph(Shefchek et al., 2020) or KG-Hub(Harry Caufield, Putman, et al., 2023).

## Further research is required on narrative outputs of LLM-based gene set summarization

Our methods and study focused on using LLM-based methods to generate GO-term-based summaries of experiments from underlying gene sets, analogous to standard term enrichment. We also demonstrated the ability to create narrative summaries of these gene sets, and even to provide mechanistic explanations of underlying biological processes. However, we did not attempt to evaluate this content, beyond demonstrating that this also frequently changed from run to run. Fully evaluating narrative output is much more challenging, as such an evaluation would be subjective and would itself require NLP techniques to automate, with attendant dangers of circularity.

However, while we were not able to systematically evaluate the quality of these narratives, we found many to sound plausible and even compelling. It is important to study this further, as there may be a temptation to use LLMs to “tell stories” about data. This could be risky and highly problematic, due to well-known issues such as hallucinations and bias. We were not able to detect hallucination and bias in the term summarization tasks, but it is important to note that this is a highly constrained task with strong in-context cues; even here we are unable to guarantee the absence of these problems, and we were easily able to induce hallucinations by asking for something the model is unable to deliver (computed p-values).

When we move from summarization in the form of controlled term lists to more open-ended summarization tasks, such as generating narrative summaries, the dangers increase. These may be more nuanced than outright confabulation of results. LLMs have been shown to exhibit “behavior” such as sycophancy (telling the user what they want to hear) and sandbagging (detecting naivety on the part of the user and providing false information)(Bowman, 2023), all of which are potential risks when interpreting scientific data using background knowledge.

Although our evaluation ignored the textual summaries when parsing term lists, we noticed one occasion when the prompt completion provided additional misleading commentary at the end:

*Note: These terms were statistically over-represented among the listed genes. The cytoskeletal reorganization was not statistically significant enough to be included. The underlying biological mechanism is likely related to the regulation of intracellular trafficking and signaling pathways, which are important for the maintenance of cellular homeostasis*.

We know for a fact that the LLM did not perform a statistical test, despite what it may be telling us. However, it is true that cytoskeletal reorganization (closest match in GO is GO:0007010, “cytoskeletal organization”) is the function of some of the genes, but not enough to reach the level of statistical significance. The text above is therefore partially correct, but only by accident. However, results like this can easily ‘sandbag’ a researcher into over-interpreting or believing incorrect interpretations.

Our results are complementary to yet consistent with the evaluation performed by Hu et al. (Hu et al., 2023), in which they used GPT-4 to generate a single label for each gene set. This evaluation was performed on (a) gene sets corresponding to existing GO terms, checking whether identical or similar labels are recapitulated (b) gene sets from omics data, using manual evaluation by experts. In both cases, no additional contextual information about gene function was passed as context into the prompt, thus this corresponds most closely to the “no synopsis” experiments we performed. Our experiments included both omics gene sets and gene sets from existing GO terms, and included additional random perturbations. A key difference is that Hu et al. examine whether a single unique descriptive label can be generated from a gene set, giving a single uniform picture of gene function, whereas we generated lists of labels, and grounded these to GO terms, more similar to standard gene enrichment. While these are exploring complementary aspects and have different evaluation strategies, our results are consistent in that the results were largely free of hallucinations. Both studies also underpin the need for standard methods for evaluating the results of gene set enrichment and summarization methods.

We note that statistical approaches for standard gene set enrichment analysis also suffer from drawbacks that may be complemented by approaches such as with LLMs. For example, statistical methods can have sample size dependent effects and minimal numbers of instances required for accurate statistical estimates. These approaches are also subject to biases from the input or reference data used in the enrichment analysis with many types of bias corrections attempted but as a result introducing differences that can render results incompatible. Other more operational issues around data integration, uneven data representation across species and data types, as well as the current state of reference knowledge resources compared to what is known in the literature, are other challenges associated with statistical enrichment methods. In theory, LLMs with billions of parameters trained on a wide range of input knowledge, potentially fine tuned for performance on dedicated tasks, may be help overcome some of these issues in the context of a single unified reference model.

This danger is compounded when we consider the fact that the leading models used in the kinds of higher-order instruction-based prompting demonstrated in this paper are not open, with essentially inscrutable training data(Bender et al., 2021), much of it derived from massive numbers of websites, likely including datasets such as the Colossal Clean Crawled Corpus (C4). Despite being cleaned, the C4 still includes significant content from websites favoring white supremacist thought(Dodge et al., 2021; “Inside the Secret List of Websites That Make AI like ChatGPT Sound Smart,” n.d.). The prospect of using models trained on this content to interpret human genetics data, bypassing human involvement, should be alarming.

Some of these dangers can be mitigated by moving towards open models where training sets are transparent, and by using curated trusted KB content via in-context cues. However, even with these measures, scientific interpretations derived from current LLMs should not be used in place of standard KB enrichment systems.

# Conclusions

We investigated the ability of LLMs to perform gene set function summarization as compared to standard ontology-based gene set enrichment analyses. We compared different models and different sources of gene descriptions, and found that while the oldest models performed most poorly, the newer GPT-4 did not substantially outperform GPT-3.5 albeit using different input data (GO term summaries and gene names, respectively). Thus considering the latest model, when token length limits are controlled for, using precise ontological descriptors derived from high quality manual curation and vetted propagation methods outperforms either narrative descriptions or relying on the models’ latent KB. When compared against standard enrichment, the LLM-generated results are typically plausible, relevant, and largely free of hallucination.

However, the most precise and informative term is usually missed, likely reflecting the lack of an ability to generalize. The LLM approach also lacks statistical rigor, and the model is unable to natively provide p-values or reliable quantitative indicators of relevance of terms. Additionally, performance varies when genes are less well known, especially in the case of model organism genes. Results are also highly non-deterministic, with different terms found on different runs.

Nevertheless, the results are impressive given the relative newness of instruction-based LLMs, and illustrate powerful textual manipulation and in-context capabilities. The ability to generate narrative summaries alongside term lists is compelling; however, there are substantial risks of hallucination and bias associated with this approach. Our results underscore the need for high quality up to date human-curated KBs to assist with the interpretation of scientific data.

# Acknowledgements

We thank Val Wood for constructive comments on the manuscript.

## Funding

This work was supported by the National Institutes of Health National Human Genome Research Institute [HG010860, HG012212]; National Institutes of Health Office of the Director [R24 OD011883]; and the Director, Office of Science, Office of Basic Energy Sciences, of the US Department of Energy [DE-AC0205CH11231 to J.H.C., H.H., N.L.H., M.J., S.M., J.T.R, and C.J.M.]. We also gratefully acknowledge Bosch Research for their support of this research project.

# References

Alliance of Genome Resources Consortium. (2020). Alliance of Genome Resources Portal: unified model organism research platform. *Nucleic Acids Research*, *48*(D1), D650–D658.

Baker, E., Bubier, J. A., Reynolds, T., Langston, M. A., & Chesler, E. J. (2016). GeneWeaver: data driven alignment of cross-species genomics in biology and disease. *Nucleic Acids Research*, *44*(D1), D555–D559.

Ballouz, S., Pavlidis, P., & Gillis, J. (2017). Using predictive specificity to determine when gene set analysis is biologically meaningful. *Nucleic Acids Research*, *45*(4), e20.

Bastian, F. B., Roux, J., Niknejad, A., Comte, A., Fonseca Costa, S. S., de Farias, T. M., Moretti, S., Parmentier, G., de Laval, V. R., Rosikiewicz, M., Wollbrett, J., Echchiki, A., Escoriza, A.,

Gharib, W. H., Gonzales-Porta, M., Jarosz, Y., Laurenczy, B., Moret, P., Person, E., … Robinson-Rechavi, M. (2021). The Bgee suite: integrated curated expression atlas and comparative transcriptomics in animals. *Nucleic Acids Research*, *49*(D1), D831–D847.

Bauer, S., Gagneur, J., & Robinson, P. N. (2010). GOing Bayesian: model-based gene set analysis of genome-scale data. *Nucleic Acids Research*, *38*(11), 3523–3532.

Bender, E. M., Gebru, T., McMillan-Major, A., & Shmitchell, S. (2021). On the Dangers of Stochastic Parrots: Can Language Models Be Too Big? 🦜. *Proceedings of the 2021 ACM Conference on Fairness, Accountability, and Transparency*, 610–623.

Bombieri, M., Fiorini, P., Ponzetto, S. P., & Rospocher, M. (2024). Do LLMs Dream of Ontologies? In *arXiv [cs.CL]*. arXiv. http://arxiv.org/abs/2401.14931

Bowman, S. R. (2023). Eight things to know about large language models. In *arXiv [cs.CL]*. arXiv. http://arxiv.org/abs/2304.00612

Brown, T. B., Mann, B., Ryder, N., Subbiah, M., Kaplan, J., Dhariwal, P., Neelakantan, A., Shyam, P., Sastry, G., Askell, A., Agarwal, S., Herbert-Voss, A., Krueger, G., Henighan, T., Child, R., Ramesh, A., Ziegler, D. M., Wu, J., Winter, C., … Amodei, D. (2020). Language Models are Few-Shot Learners. In *arXiv [cs.CL]*. arXiv. http://arxiv.org/abs/2005.14165

Creators Chris Mungall1 Harshad1 Patrick Kalita1 Charles Tapley Hoyt2 Sujay Patil1 marcin p. joachimiak1 Joe Flack3 David Linke4 Nomi Harris1 Sierra Moxon5 Kim Rutherford6 Nico Matentzoglu7 Deepak8 Harry Caufield1 Vinícius de Souza Glass9 Jules Jacobsen10 Justin Reese11 Manuel Lera Ramirez Shawn Tan Show affiliations 1. Lawrence Berkeley National Laboratory 2. Harvard Medical School 3. @jhu-bids 4. Leibniz-Institut für Katalyse e. V. (LIKAT) 5. LBNL 6. Uni of Cambridge / @PomBase 7. semanticly Ltd 8. SIB Swiss Institute of Bioinformatics 9. @det-lab @tis-lab @monarch-initiative 10. Queen Mary University of London 11. Lawrence Berkeley National Lab. (2023). *INCATools/ontology-access-kit: v0.5.6*. https://doi.org/10.5281/zenodo.7938070

Devlin, J., Chang, M.-W., Lee, K., & Toutanova, K. (2018). BERT: Pre-training of Deep

Bidirectional Transformers for Language Understanding. In *arXiv [cs.CL]*. arXiv. http://arxiv.org/abs/1810.04805

Dodge, J., Sap, M., Marasovic, A., Agnew, W., & Ilharco, G. (2021). Documenting the english colossal clean crawled corpus. *arXiv Preprint arXiv*.

Dolgalev, I. (n.d.). msigdbr: MSigDB gene sets for multiple organisms in a tidy data format. *R Package Version*.

Duck, G., Nenadic, G., Filannino, M., Brass, A., Robertson, D. L., & Stevens, R. (2016). A Survey of Bioinformatics Database and Software Usage through Mining the Literature. *PloS One*, *11*(6), e0157989.

Fabregat, A., Jupe, S., Matthews, L., Sidiropoulos, K., Gillespie, M., Garapati, P., Haw, R., Jassal, B., Korninger, F., May, B., Milacic, M., Roca, C. D., Rothfels, K., Sevilla, C., Shamovsky, V., Shorser, S., Varusai, T., Viteri, G., Weiser, J., … D'Eustachio, P. (2018). The Reactome Pathway Knowledgebase. *Nucleic Acids Research*, *46*(D1), D649–D655.

Garcia, F. J., Sun, N., Lee, H., Godlewski, B., Mathys, H., Galani, K., Zhou, B., Jiang, X., Ng, A. P., Mantero, J., Tsai, L.-H., Bennett, D. A., Sahin, M., Kellis, M., & Heiman, M. (2022). Single-cell dissection of the human brain vasculature. *Nature*, *603*(7903), 893–899.

Gene Ontology Consortium. (2023). The Gene Ontology Knowledgebase in 2023. *Genetics*. https://doi.org/10.1093/genetics/iyad031

Green, M. L., & Karp, P. D. (2006). The outcomes of pathway database computations depend on pathway ontology. *Nucleic Acids Research*, *34*(13), 3687–3697.

Harry Caufield, J., Hegde, H., Emonet, V., Harris, N. L., Joachimiak, M. P., Matentzoglu, N., Kim, H., Moxon, S. A. T., Reese, J. T., Haendel, M. A., Robinson, P. N., & Mungall, C. J. (2023). Structured prompt interrogation and recursive extraction of semantics (SPIRES): A method for populating knowledge bases using zero-shot learning. In *arXiv [cs.AI]*. arXiv. http://arxiv.org/abs/2304.02711

Harry Caufield, J., Putman, T., Schaper, K., Unni, D. R., Hegde, H., Callahan, T. J., Cappelletti,

L., Moxon, S. A. T., Ravanmehr, V., Carbon, S., Chan, L. E., Cortes, K., Shefchek, K. A., Elsarboukh, G., Balhoff, J. P., Fontana, T., Matentzoglu, N., Bruskiewich, R. M., Thessen, A. E., … Reese, J. T. (2023). KG-Hub -- Building and Exchanging Biological Knowledge Graphs. In *arXiv [q-bio.QM]*. arXiv. http://arxiv.org/abs/2302.10800

Hou, W., & Ji, Z. (2023). Reference-free and cost-effective automated cell type annotation with GPT-4 in single-cell RNA-seq analysis. *bioRxiv : The Preprint Server for Biology*. https://doi.org/10.1101/2023.04.16.537094

Hu, M., Alkhairy, S., Lee, I., Pillich, R. T., Bachelder, R., Ideker, T., & Pratt, D. (2023). Evaluation of large language models for discovery of gene set function. *ArXiv*. https://doi.org/10.1038/npre.2010.5025.1

Hung, J.-H., Yang, T.-H., Hu, Z., Weng, Z., & DeLisi, C. (2012). Gene set enrichment analysis: performance evaluation and usage guidelines. *Briefings in Bioinformatics*, *13*(3), 281–291.

Inside the secret list of websites that make AI like ChatGPT sound smart. (n.d.). *The Washington Post*. Retrieved May 7, 2023, from https://www.washingtonpost.com/technology/interactive/2023/ai-chatbot-learning/

Jin, Q., Yang, Y., Chen, Q., & Lu, Z. (2023). GeneGPT: Augmenting Large Language Models with Domain Tools for Improved Access to Biomedical Information. *ArXiv*. https://www.ncbi.nlm.nih.gov/pubmed/37131884

Ji, Z., Lee, N., Frieske, R., Yu, T., Su, D., Xu, Y., Ishii, E., Bang, Y., Dai, W., Madotto, A., & Fung, P. (2022). Survey of Hallucination in Natural Language Generation. In *arXiv [cs.CL]*. arXiv. http://arxiv.org/abs/2202.03629

Joachimiak, M. p. (2023). *monarch-initiative/enrichgpt-results: initial*. https://doi.org/10.5281/zenodo.7893613

Karp, P. D. (2016). How much does curation cost? *Database: The Journal of Biological Databases and Curation*, *2016*. https://doi.org/10.1093/database/baw110

Kishore, R., Arnaboldi, V., Van Slyke, C. E., Chan, J., Nash, R. S., Urbano, J. M., Dolan, M. E.,

Engel, S. R., Shimoyama, M., Sternberg, P. W., & Genome Resources, T. A. O. (2020). Automated generation of gene summaries at the Alliance of Genome Resources. *Database: The Journal of Biological Databases and Curation*, *2020*. https://doi.org/10.1093/database/baaa037

Köhler, S., Gargano, M., Matentzoglu, N., Carmody, L. C., Lewis-Smith, D., Vasilevsky, N. A., Danis, D., Balagura, G., Baynam, G., Brower, A. M., Callahan, T. J., Chute, C. G., Est, J. L., Galer, P. D., Ganesan, S., Griese, M., Haimel, M., Pazmandi, J., Hanauer, M., … Robinson, P. N. (2021). The Human Phenotype Ontology in 2021. *Nucleic Acids Research*, *49*(D1), D1207–D1217.

Lee, J., Yoon, W., Kim, S., Kim, D., Kim, S., So, C. H., & Kang, J. (2019). BioBERT: a pre-trained biomedical language representation model for biomedical text mining. In *arXiv [cs.CL]*. arXiv. https://doi.org/10.1093/database/bay073/5055578

Lopriore, P., Ricciarini, V., Siciliano, G., Mancuso, M., & Montano, V. (2022). Mitochondrial Ataxias: Molecular Classification and Clinical Heterogeneity. *Neurology International*, *14*(2), 337–356.

Mi, H., Muruganujan, A., Ebert, D., Huang, X., & Thomas, P. D. (2019). PANTHER version 14: more genomes, a new PANTHER GO-slim and improvements in enrichment analysis tools. *Nucleic Acids Research*, *47*(D1), D419–D426.

O'Leary, N. A., Wright, M. W., Brister, J. R., Ciufo, S., Haddad, D., McVeigh, R., Rajput, B., Robbertse, B., Smith-White, B., Ako-Adjei, D., Astashyn, A., Badretdin, A., Bao, Y., Blinkova, O., Brover, V., Chetvernin, V., Choi, J., Cox, E., Ermolaeva, O., … Pruitt, K. D. (2016). Reference sequence (RefSeq) database at NCBI: current status, taxonomic expansion, and functional annotation. *Nucleic Acids Research*, *44*(D1), D733–D745.

Pal, A., Umapathi, L. K., & Sankarasubbu, M. (2023). Med-HALT: Medical Domain Hallucination Test for Large Language Models. In *arXiv [cs.CL]*. arXiv. http://arxiv.org/abs/2307.15343

Putman, T. E., Schaper, K., Matentzoglu, N., Rubinetti, V. P., Alquaddoomi, F. S., Cox, C.,

Caufield, J. H., Elsarboukh, G., Gehrke, S., Hegde, H., Reese, J. T., Braun, I., Bruskiewich, R. M., Cappelletti, L., Carbon, S., Caron, A. R., Chan, L. E., Chute, C. G., Cortes, K. G., … Munoz-Torres, M. C. (2024). The Monarch Initiative in 2024: an analytic platform integrating phenotypes, genes and diseases across species. *Nucleic Acids Research*, *52*(D1), D938–D949.

Reimand, J., Isserlin, R., Voisin, V., Kucera, M., Tannus-Lopes, C., Rostamianfar, A., Wadi, L., Meyer, M., Wong, J., Xu, C., Merico, D., & Bader, G. D. (2019). Pathway enrichment analysis and visualization of omics data using g:Profiler, GSEA, Cytoscape and EnrichmentMap. *Nature Protocols*, *14*(2), 482–517.

Ronacher, A. (2008). Jinja2 documentation. *Welcome to Jinja2—Jinja2 Documentation (2. 8-Dev)*. http://mitsuhiko.pocoo.org/jinja2docs/Jinja2.pdf

Shefchek, K. A., Harris, N. L., Gargano, M., Matentzoglu, N., Unni, D., Brush, M., Keith, D., Conlin, T., Vasilevsky, N., Zhang, X. A., Balhoff, J. P., Babb, L., Bello, S. M., Blau, H., Bradford, Y., Carbon, S., Carmody, L., Chan, L. E., Cipriani, V., … Osumi-Sutherland, D. (2020). The Monarch Initiative in 2019: an integrative data and analytic platform connecting phenotypes to genotypes across species. *Nucleic Acids Research*, *48*(D1), D704–D715.

Tomczak, A., Mortensen, J. M., Winnenburg, R., Liu, C., Alessi, D. T., Swamy, V., Vallania, F., Lofgren, S., Haynes, W., Shah, N. H., Musen, M. A., & Khatri, P. (2018). Interpretation of biological experiments changes with evolution of the Gene Ontology and its annotations. *Scientific Reports*, *8*(1), 5115.

Toufiq, M., Rinchai, D., Bettacchioli, E., Kabeer, B. S. A., Khan, T., Subba, B., White, O., Yurieva, M., George, J., Jourde-Chiche, N., Chiche, L., Palucka, K., & Chaussabel, D. (2023). Harnessing large language models (LLMs) for candidate gene prioritization and selection. *Journal of Translational Medicine*, *21*(1), 728.

Touvron, H., Martin, L., Stone, K., Albert, P., Almahairi, A., Babaei, Y., Bashlykov, N., Batra, S., Bhargava, P., Bhosale, S., Bikel, D., Blecher, L., Ferrer, C. C., Chen, M., Cucurull, G.,

Esiobu, D., Fernandes, J., Fu, J., Fu, W., … Scialom, T. (2023). Llama 2: Open Foundation and Fine-Tuned Chat Models. In *arXiv [cs.CL]*. arXiv. http://arxiv.org/abs/2307.09288

Wadi, L., Meyer, M., Weiser, J., Stein, L. D., & Reimand, J. (2016). Impact of outdated gene annotations on pathway enrichment analysis. *Nature Methods*, *13*(9), 705–706.

Wang, Q., Armenia, J., Zhang, C., Penson, A. V., Reznik, E., Zhang, L., Minet, T., Ochoa, A., Gross, B. E., Iacobuzio-Donahue, C. A., Betel, D., Taylor, B. S., Gao, J., & Schultz, N. (2018). Unifying cancer and normal RNA sequencing data from different sources. *Scientific Data*, *5*, 180061.

Wei, J., Tay, Y., Bommasani, R., Raffel, C., Zoph, B., Borgeaud, S., Yogatama, D., Bosma, M., Zhou, D., Metzler, D., Chi, E. H., Hashimoto, T., Vinyals, O., Liang, P., Dean, J., & Fedus, W. (2022). Emergent Abilities of Large Language Models. In *Transactions on Machine Learning Research*. https://openreview.net/pdf?id=yzkSU5zdwD

Zhao, K., & Rhee, S. Y. (2023). Interpreting omics data with pathway enrichment analysis. *Trends in Genetics: TIG*, *39*(4), 308–319.